# Molecular docking and binding mode analysis of selected FDA approved drugs against COVID-19 selected key protein targets: An effort towards drug repurposing to identify the combination therapy to combat COVID-19

Atanu Barik#, Geeta Rai,‖ Gyan Modi #*

#Department of Pharmaceutical Engineering & Technology, Indian Institute of Technology (Banaras Hindu University), Varanasi, 221005, India

‖Department of Molecular and Human Genetics, Institute of Science, Banaras Hindu University, Varanasi, 221005, India

## Abstract

The emergence of COVID-19 has severely compromised the arsenal of antiviral and antibiotic drugs. Drug discovery is a multistep process with a high failure rate, high cost and it takes approximately 10-12 years for the development of new molecules into the clinical candidate. On the other side, drug repurposing also called old drugs for new uses, is an attractive alternative approach for a new application of marketed FDA approved or investigational drugs. In the current pandemic situation raised due to COVID-19, repurposing of existing FDA approved drugs are emerging as the first line of the treatment. The causative viral agent of this highly contagious disease and acute respiratory syndrome coronavirus (SARS-CoV) shares high nucleotide similarity. Therefore, many existing viral targets are structurally expected to be similar to SARS-CoV and likely to be inhibited by the same compounds. Here, we selected three viral key proteins based on their vital role in viral life cycle: ACE2 (helps in entry into the human host), viral nonstructural proteins RNA-dependent RNA polymerase (RdRp) NSP12, and NSP16 which helps in replication, and viral latency (invasion from immunity). The FDA approved drugs chloroquine (**CQ**), hydroxychloroquine (**HCQ**), remdesivir **(RDV)** and arbidol **(ABD)** are emerging as promising agents to combat COVID-19. Our hypothesis behind the docking studies is to determine the binding affinities of these drugs and identify the key amino acid residues playing a key role in their mechanism of action. The docking studies were carried out through Autodock and online COVID-19 docking server. Further studies on a broad range of FDA approved drugs including few more protein targets, molecular dynamics studies, *in-vitro* and *in-vivo* biological evaluation are required to identify the combination therapy targeting various stages of the viral life cycle.

## Introduction

The recent outbreak of the coronavirus disease 2019 (COVID-19) caused by the new coronavirus 2019-nCoV. Although the original epicenter of the COVID-19 outbreak in December 2019 was located in the seafood market of Wuhan city, Hubei province of China. [1] The disease has spread to more than 190 countries with over 1,353,361 confirmed cases and over 79,235 confirmed deaths worldwide in a very short span of approximately three months as of April 08, 2020. [2] The extremely rapid chain transmission and deadly pathogenic nature of this virus are some of the major concerns for all sector of society across the world.[3] So far, the best possible solution of this nightmare viral disease is only social distancing which is resulting in the mandatory isolations/quarantines and lockdown across the word at different time points. [4] Further, the lives of the millions of the people and the world economy have been severely impacted due to this outbreak. The pathogenic nature of this agent implicates a plausible sever biothreat across the globe. The severity of the COVID-19 outbreak could force to impose major changes to health system across the globe and possible bring a major challenges to human being for the survival followed struggling to keep the global economy on correct pace, if the spread of the virus is not effectively controlled. [5, 6]

The causative agent of this devastating disease COVID-19 belongs to Beta coronavirus which shares 89.1% nucleotide similarity with acute respiratory syndrome coronavirus (SARS-CoV). [7] The key amino acid residues are conserved in many viral key drug targets, including those found in both SARS-CoV and COVID-19 pathogens. Consequently, many common viral targets are structurally similar to SARS-CoV and likely to be inhibited by the same compounds. The emergence of COVID-19 has severely compromised the arsenal of antiviral and antibiotic drugs. New compounds and targets are needed to meet the growing threat from Beta coronavirus. The possibility that COVID-19 and other viruses/bacteria can be perniciously engineered as biowarfare agents creates another demand for new treatment options. Drug discovery is a multistep process with a high attrition rate, substantial cost and slow pace of development of new molecules into the clinical candidate. On the other side, drug repurposing also called old drugs for new uses, is an attractive alternative approach for the new application of FDA approved or investigational drugs that are outside the scope of the original medical condition. [8] There is a plethora of literature evidence indicated the vital application of drug repurposing for various infectious diseases.[9-11] Given the current scenario of COVID-19 outbreak across the world and the complex nature of the disease repurposing of existing FDA approved drugs is the first line of the treatment.

Given the pandemic nature of the disease, there is a hard-pressing need to uncover the possible treatment/s as early as possible. The researchers and physicians have been putting sincere efforts to understand this new virus, the pathophysiology of the disease to and the possible therapeutic effective agents and vaccines. Tremendous effects have been done to study the newly emerged virus and find potent drugs for clinical usage. A PubMed search with key work COVID-19 resulted in more than 900 publications from November 01, 2019, to April 31, 2020. A glance at the title of these papers indicated that the majority are focused on manifestations and treatment options.

It has been shown in the literature that the FDA approved antimalarial and autoimmune disease drug **chloroquine (CQ)** and **hydroxychloroquine (HCQ)**, known to inhibit viral infection by raising endosomal pH necessary for the interaction of virus/cell fusion.[12, 13] [14] **CQ and HCQ** interferes ACE2 glycosylation or glycosyltransferases within human cells or inhibits sialic acid biosynthesis by inhibiting quinone reductase 2. The immunomodulatory property of chloroquine synergistically involved with the anti-viral property. The antiviral drug **remdesivir (RDV)** can inhibit the growth of COVID-19 and found to be efficacious in the clinic in combination.[13, 15, 16] [17]The mechanism of action of **RDV** is under investigation, however, the high sequence similarity (>95%) between COVID-19 and SARS-CoV RNA-dependent RNA polymerase (RdRp) indicating the inhibition of RdRp. Similarly, the broad-spectrum virus-host cell fusion inhibitor marketed antiviral drug **arbidol (ABD)**, prevents the entry of the virus entry into the host cell acting through ACE2 has entered into a clinical trial for the treatment of COVID-19. [18, 19]

Inspired from the current pandemic situation and our experience in drug repurposing/drug design through computational and medicinal chemistry tools, we have carried docking studies with selected FDA approved drugs against COVID-19 selected key protein targets. Given the complex nature of this rapidly and severely infecting lower respiratory system disease, our hypothesis is to identify FDA approved drug combinations targeting the different key viral proteins. The selection of the key proteins is based on the virus life cycle starting from attachment onto the human host, replication, viral latency (invasion from immunity), release and complex nature of the disease. The selection of four key targets: **PDB ID**: **6LZG** which is the structure of novel coronavirus spike receptor-binding domain complexed with its receptor ACE2; **PDB ID**: **6NUR** a recently identified SARS-Coronavirus NSP12 bound to NSP7 and NSP8 co-factors, **PDB ID**: **6W4H** the crystal structure of NSP16-NSP10 Complex from SARS-CoV-2 of SARS-COV, and homology model of nonstructural protein 12 (nsp12) RNA-dependent RNA polymerase (RdRp) with RNA and without RNA. While we carried out the docking with the homology model of RdRP, the crystal structure of **PDB ID: 6M71** SARS-Cov-2 RNA-dependent RNA polymerase in complex with cofactors was reported. So, we carried out the docking using this crystal structure as well. The online COVID-19 Docking Server" (http://ncov.schanglab.org.cn) which contains the homology model of COVID-19 RdRp with RNA and without RNA was taken in the experiment. The non-structural proteins, nsp16-nsp10 and 3plPro play an important role in virus genome replication and evasion from innate immunity.

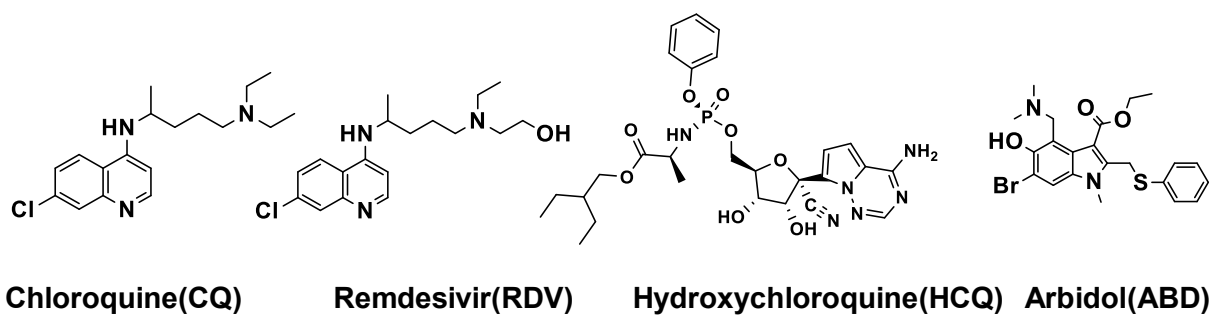

**Chloroquine(CQ)**     **Remdesivir(RDV)**     **Hydroxychloroquine(HCQ)**   **Arbidol(ABD)**

Figure 1: Chemical structures of the selected FDA approved drugs for docking studies

**Experimental section**

**Docking methodology:** Molecular docking studies are a method of providing valuable information on the rationale of designing ligands for a particularly active site of a well-known macromolecule. This is an economic and modern trend of drug discovery where technology base ligand-protein interaction reveals the pre-synthesizing possibilities. The *in-silico* study of the four FDA drugs chloroquine (CQ), hydroxychloroquine (HCQ), remdesivir (RDV), and arbidol (ABD) (**Figure 1**) was performing blind docking both in online and offline modes.

The offline docking was carried out using Autodock 4.2 program package (http://autodock.scripps.edu/).[20] The X-ray crystal structures of three different proteins were retrieved from the RCSB protein data bank (www.rscb.org). Macromolecule having PDB ID: 6LZG which is the structure of novel coronavirus spike receptor-binding domain complexed with its receptor ACE2; 6NUR a recently identified SARS-Coronavirus NSP12 bound to NSP7 and NSP8 co-factors, 6M71 SARS-Cov-2 RNA-dependent RNA polymerase in complex with cofactors and 6W4H the crystal Structure of NSP16 - NSP10 Complex from SARS-CoV-2 of SARS-COV were selected based on the resolution and similarity with COVID-19. All the protein and ligand preparation were performed using MGL Tools 1.5.6 and Autodock Tool (ADT). The energy was minimized using Chem 3D program file. Standard precision with flexible ligand sampling was used and no specific grid was used at any stage of the experiment. The ADT was used to calculate the binding free energies and inhibition constant of the best-docked complex of the aforementioned proteins. Observation and visualization of the results were carried out using the latest version of Discovery Studio Visualizer (https://www.3dsbiovia.com/products/collaborative-science/biovia-discovery-studio/visualization-download.php). Finally, the top binding modes are according to their energy calculation as the default parameter in AutoDoc.

Further to expand the specificity, we performed online docking by using "The COVID-19 Docking Server" (http://ncov.schanglab.org.cn) which contains the structure of the proteins involved in SARS-CoV virus life cycle based on the homologs of coronavirus. [21, 22] Ligands preparation and energy minimization were carried out using the Chem 3D application and uploaded to the server in .mol2 format. Autodock Vina was used as a docking engine in the sever and prepared Nonstructural protein 12 (nsp12) RNA-dependent RNA polymerase (RdRp) with RNA and without RNA was taken in the experiment. All the standard parameters were set as default except the exhaustiveness value is set to 12 to achieve higher accuracy. Visualization of the result was done by the latest version of Discovery Studio Visualizer and binding free energy was collected from the aforesaid web server.[23]

**Results and discussion:**

**1. The binding mode analysis and predicated binding affinity calculations of chloroquine (CQ) against surface receptor and nonstructural proteins**: The Auto dock software was used for molecular docking studies of chloroquine (CQ) against the

novel coronavirus spike receptor-binding domain complexed with its receptor ACE2 (PDB ID 6LZG) revealed the interaction of CQ  with TRP349 ALA348 PRO346 ASP350 as shown in **Figure 2 A**. The negative values of the binding free energy (-6.67 kcal/mole) further indicated the stability of the complex (**Table 1**). The receptor pocket and top 10 conformers are shown in **figure 2A**. Similarly, the docking studies and binding mode analysis of CQ against SARS-Coronavirus NSP12 bound to NSP7 and NSP8 co-factors (PDB ID 6NUR) showed the interactions with MET542 ALA558 ARG624 SER682 SER681 ASP623 ASP452 ALA554 (**figure 2B**). The negative free energy calculation (-6.49 kcal/mole) as shown in **table 1** is an indication of the interaction of CQ with nonstructural viral proteins. Our docking studies of CQ with NSP16- NSP10, however, our docking methodology did not work very well with this target resulted in the high positive values of binding energy (data are not shown here). We will further improve our next publications.

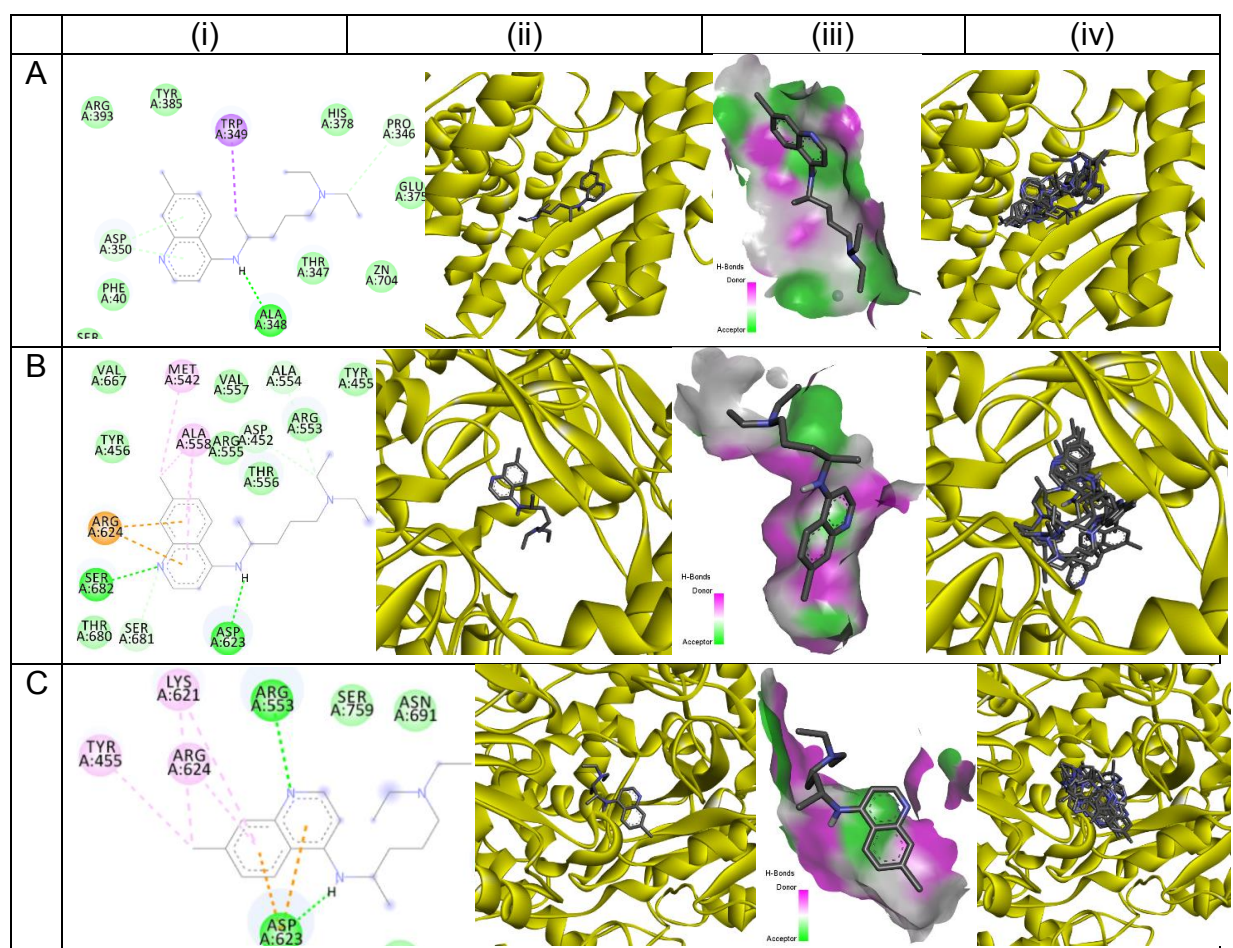

**Figure 2**: Conformational changes observed due to the binding of ligand Chloroquine with **A**. PDB ID: 6LZG, **B**. PDB ID: 6NUR, **C.** PDB ID: 6M71; left to right best pose, interaction 2D image of best pose, receptor pocket, and top10 conformers.

**Table 1:** Binding affinity of chloroquine ($C_{18}H_{26}ClN_3$) with the targets PDB ID: 6LZG, 6NUR, and 6M71

| Rank | PDB ID: 6LZG | | | PDB ID: 6NUR | | | PDB ID: 6M71 | | |
|---|---|---|---|---|---|---|---|---|---|
| | Free Energy of Binding (kcal/mol) | Predicted Inhibition Constant (µM) | Interacting Amino Acids | Free Energy of Binding (kcal/mol) | Predicted Inhibition Constant (µM) | Interacting Amino Acids | Free Energy of Binding (kcal/mol) | Predicted Inhibition Constant (µM) | Interacting Amino Acids |
| 1 | -6.67 | 12.94 | TRP349 ALA348 PRO346 ASP350 | -6.49 | 17.52 | MET542 ALA558 ARG624 SER682 SER681 ASP623 ASP452 ALA554 | -3.44 | 3.03 | ARG553 LYS621 ARG624 TYR455 ASP623 |
| 2 | -6.41 | 19.93 | ARG393 ASP350 ASP382 ALA348 | -5.81 | 55.13 | ARG555 ARG624 ASP623 | -3.27 | 4.03 | LYS621 ASP623 ARG553 |
| 3 | -6.35 | 22.3 | ARG393 ASP382 HIS401 GLU375 ASP350 | -5.60 | 78.54 | ASP623 LYS621 ARG624 TYR455 ARG553 | -3.25 | 4.15 | ARG553 LYS621 ASP623 ARG624 TYR455 CYS622 |
| 4 | -6.20 | 28.47 | HIS378 TRP349 ALA348 ASP350 | -5.27 | 138.12 | LYS545 ARG555 ARG553 THR556 ASP623 ASP452 | -3.24 | 4.21 | ASP760 ASP623 TYR455 LYS621 ARG553 |
| 5 | -6.11 | 32.98 | HIS378 HIS345 ALA348 Zn704 | -5.12 | 175.15 | ASP623 SER682 ALA558 VAL557 ASP760 ARG624 MET542 SER681 | -3.11 | 5.24 | LYS621 ASP623 ARG624 TYR455 ASP760 |
| 6 | -6.06 | 36.02 | HIS345 HIS378 ALA348 ASP382 | -5.06 | 194.17 | THR556 ASP623 ARG555 LYS545 ASP760 ARG553 ASP452 | -3.01 | 6.17 | ASN691 ASP760 LYS621 ARG624 TYR455 ARG553 ASP623 |
| 7 | -6.04 | 37.31 | ALA348 HIS401 ARG393 ASP350 ASP382 GLU375 PRO346 | -5.01 | 212.38 | ALA558 THR680 ARG624 ASP623 THR556 | -2.73 | 9.95 | TYR455 LYS621 ARG553 ASP760 |
| 8 | -6.03 | 38.28 | TYR385 HIS378 ALA348 | -4.86 | 272.55 | VAL557 ALA558 THR556 TYR455 TYR456 VAL667 SER681 MET542 ARG624 ASP452 | -2.71 | 10.26 | TYR455 LYS621 ARG624 ARG553 ASP623 |
| 9 | -5.84 | 52.11 | HIS378 ALA348 PRO346 ASP382 | -4.86 | 272.16 | LYS621 ASP623 ASP760 TYR455 ARG553 | -2.68 | 10.92 | ARG553 ASP623 LYS621 CYS622 TYR619 |
| 10 | -5.68 | 68.86 | HIS374 ASP382 HIS378 HIS401 ALA348 GLU375 | -4.76 | 324.02 | ALA558 VAL667 MET542 THR556 ARG624 ASP623 SER681 | -2.58 | 12.87 | ALA558 ARG624 MET542 ASP623 TYR456 |

**2. The binding mode analysis and predicated binding affinity calculations of chloroquine against the homology model of COVID-19 RNA-dependent RNA polymerase (RdRP)**: The viral nonstructural protein 12 (nsp12) or RNA-dependent RNA polymerase, RdRp) is essential for the replication and transcription of the viral genome. RdRp upon binding with the cofactors nsp7 and nsp8 catalyze the replication of RNA from an RNA template. The structure of viral RdRp of COVID-19 is not yet known, however, recently a web-based COVID-19 docking server has been launched which incorporates the homology model of RdRp built on PDB ID 6NUR, RdRp crystal structure of SARS-COV. Two structures of RdRp, one structure is constructed with RNA and others without RNA were reported for small molecules docking. Therefore, this web-based server was utilized for docking studies of CQ. The docking studies of RdRp with RNA revealed the interaction of CQ with A Chain: ALA840 ARG858 ARG555 ALA547 PHE441 ILE548 P Chain: G7 as shown in **figure 3 A**. The negative

values of the binding free energy (-7.10 kcal/mole) further indicated the stability of the complex (**Table 2**). The receptor pocket and top 10 conformers are shown in **figure 3 A**. Similarly, the docking studies of CQ on RdRp without RNA revealed the interaction of CQ with **A Chain:** ALA558 THR556 SER682 ASP452 ARG624 ASP623 as shown in **figure 3 B**. The negative values of the binding free energy (-6.20 kcal/mole) further indicated the stability of the complex (**Table 2**). The receptor pocket and top 10 conformers are shown in **figure 3 A.**

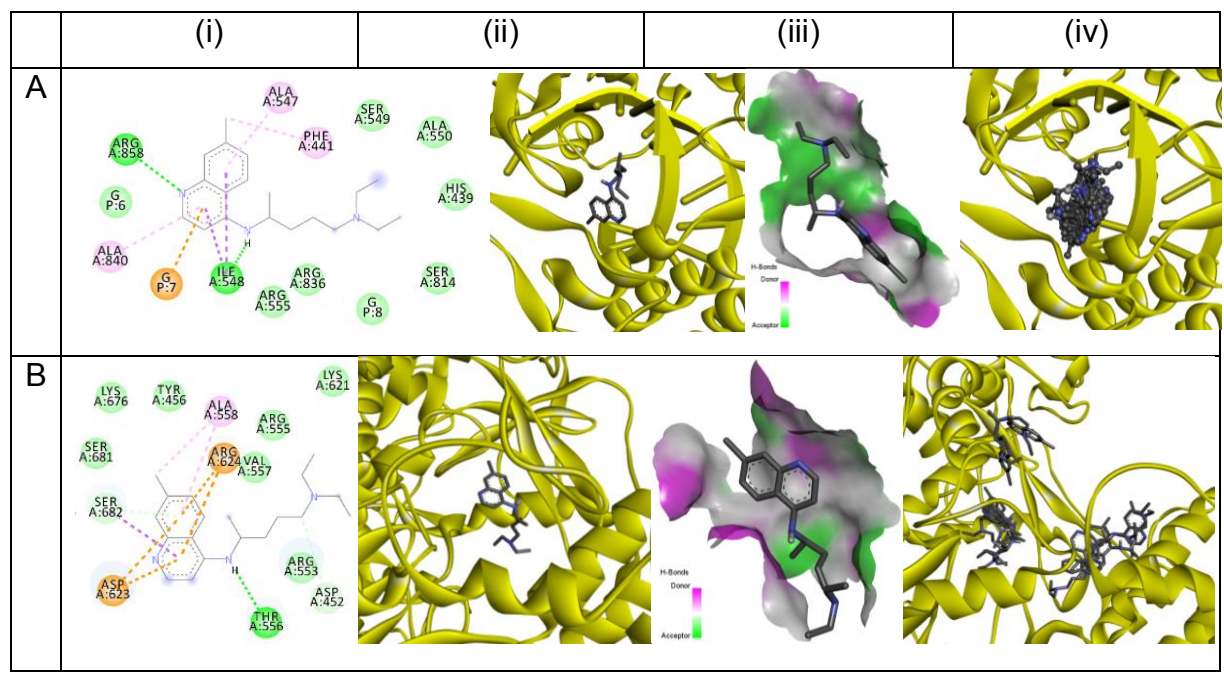

**Figure 3**: Best Conformation observed due to the binding of ligand chloroquine with Nonstructural protein 12 (nsp12) of SARS-CoV; **A:** nsp12 RdRp with RNA and **B:** nsp12 RdRp without RNA (Note: There are 2 neighboring binding pockets)

**Table 2:** Binding affinity of chloroquine ($C_{18}H_{26}ClN_3$) with the target nonstructural protein 12 (nsp12) of SARS-CoV

| Cluster Rank | NSP 12 RdRp with RNA | | NSP 12 RdRp without RNA | |
|---|---|---|---|---|
| | Energy of binding | Interacting Amino Acids | Energy of binding | Interacting Amino Acids |
| 1 | -7.10 | **A Chain:** ALA840 ARG858 ARG555 ALA547 PHE441 ILE548 <br> **P Chain:** G7 | -6.20 | **A Chain:** ALA558 THR556 SER682 ASP452 ARG624 ASP623 |
| 2 | -6.70 | **A Chain:** ARG858 ILE548 ARG836 ALA547 PHE441 <br> **P Chain:** G6 G7 | -5.90 | **A Chain:** SER682 THR556 ASP452 ASP623 |
| 3 | -6.60 | **A Chain:** ARG858 ILE548 ALA547 PHE441 <br> **P Chain:** G6 G7 | -5.80 | **A Chain:** ALA840 PHE441 ARG836 ILE548 ALA547 |
| 4 | -6.60 | **A Chain:** ALA547 LYS545 <br> **P Chain:** G6 G7 | -5.60 | **A Chain:** LYS500 ALA685 ARG569 |
| 5 | -6.30 | **A Chain:** ARG836 ILE548 ALA547 HIS439 <br> **P Chain:** G6 G7 | -5.60 | **A Chain:** SER682 MET542 ASP623 ARG624 |
| 6 | -6.30 | **A Chain:** ALA547 PHE441 ALA840 ILE548 ARG858 <br> **P Chain:** G6 G7 | -5.30 | **A Chain:** ARG836 ALA840 ILE548 |
| 7 | -6.20 | **A Chain:** ILE548 ARG836 ALA840 ARG858 <br> **P Chain:** G6 G7 | -5.20 | **A Chain:** ALA547 ILE548 ARG555 |
| 8 | -6.20 | **A Chain:** ILE548 <br> **P Chain:** C4 C5 G6 | -5.20 | **A Chain:** ARG624 ALA558 ARG555 ASP623 THR556 |
| 9 | -6.00 | **A Chain:** ALA840 PHE441 VAL844 ARG858 ILE548 <br> **P Chain:** C5 G6 | -5.20 | **A Chain:** ALA685 ALA688 TYR689 |
| 10 | -5.90 | **A Chain:** ARG555 HIS439 ILE548 ALA840 ASP845 <br> **P Chain:** G7 | -5.10 | **A Chain:** ASP845 ARG858 |

**3. The binding mode analysis and predicated binding affinity calculations of hydroxychloroquine (HCQ) against surface receptor and nonstructural proteins**: The Auto dock software was used for molecular docking studies of hydroxychloroquine (**HCQ**) against the novel coronavirus spike receptor-binding domain complexed with its receptor ACE2 (PDB ID 6LZG) revealed the interaction of **HCQ** with ==**ASP382 HIS378 HIS401 ASP350 TRP349**== as shown in **figure 4 A**. The negative values of the binding free energy (-6.65 kcal/mole) further indicated the stability of the complex (**Table 3**). The receptor pocket and top 10 conformers are shown in **figure 4A**. Similarly, the docking studies and binding mode analysis of **HCQ** against SARS-Coronavirus NSP12 bound to NSP7 and NSP8 co-factors (PDB ID 6NUR) showed the interactions with ==**ASP452**== **VAL667 MET542 ALA558 VAL557 ARG624 SER682 THR556 SER681 ASP623** (**figure 4B**). The negative free energy calculation (-6.75 kcal/mole) as shown in **table 3** is an indication of the interaction of HCQ with

nonstructural viral proteins. Our docking studies of HCQ with NSP16- NSP10, however, our docking methodology did not work very well with this target resulted in the high positive values of binding energy (data are not shown here). We will further improve our next publications.

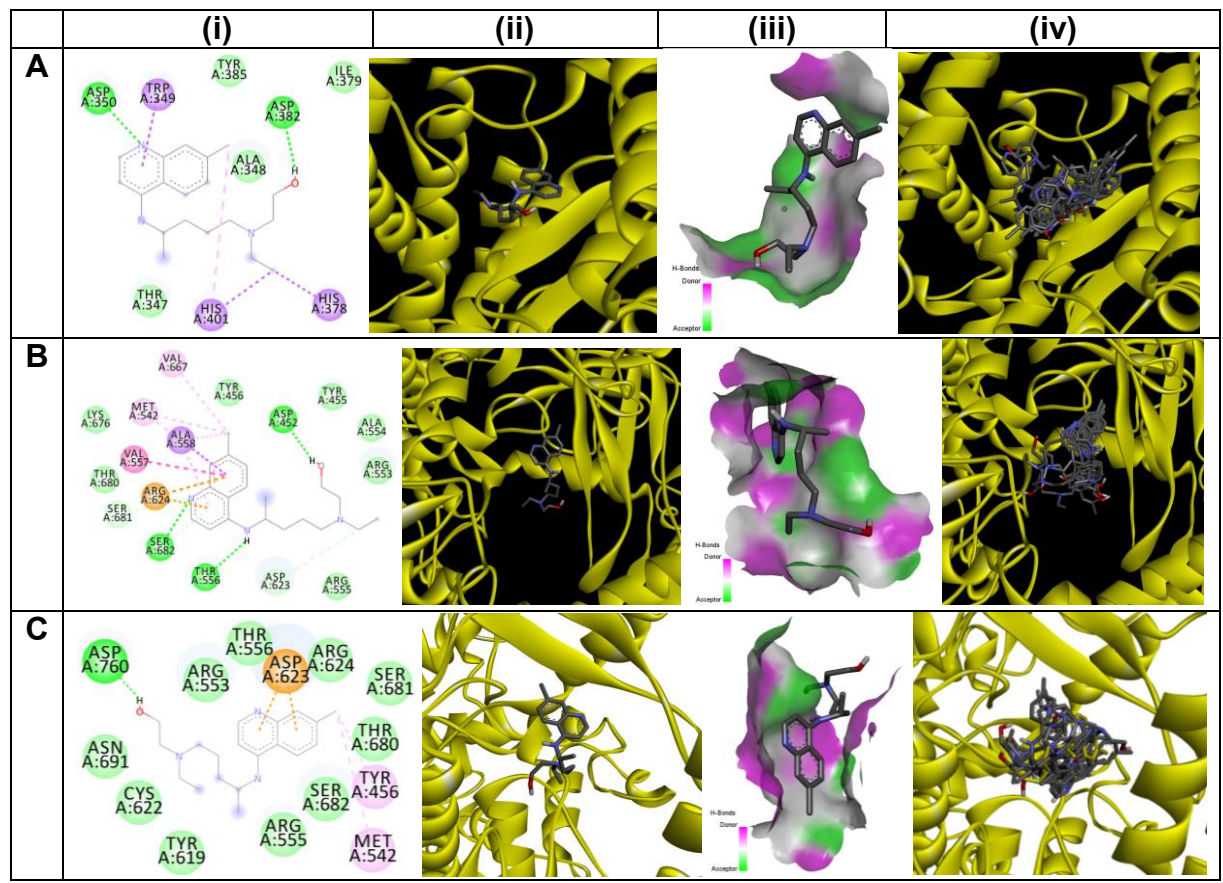

Figure 4: Conformational changes observed due to the binding of ligand Hydroxychloroquine with **A.** PDB ID: 6lzg and **B**. PDB ID: 6nur, **C.** PDB ID: 6M71; left to right there is a pose of the best conformer, interaction 2D image of best pose, receptor pocket, and top10 conformers.

**Table 3:** Binding affinity of Hydroxychloroquine ($C_{18}H_{26}OClN_3$) with the target PDB ID: 6LZG, 6NUR, and 6M71

| Rank | PDB ID: 6LZG | | | PDB ID: 6NUR | | | PDB ID: 6M71 | | |
|---|---|---|---|---|---|---|---|---|---|
| | Free Energy of Binding (kcal/mol) | Predicted Inhibition Constant (µM) | Interacting Amino Acids | Free Energy of Binding (kcal/mol) | Predicted Inhibition Constant (µM) | Interacting Amino Acids | Free Energy of Binding (kcal/mol) | Predicted Inhibition Constant (µM) | Interacting Amino Acids |
| **1** | -6.65 | 13.33 | ASP382 HIS378 HIS401 ASP350 TRP349 | -6.75 | 11.21 | ASP452 VAL667 MET542 ALA558 VAL557 ARG624 SER682 THR556 SER681 ASP623 | -3.62 | 2.22 | ASP760 ASP623 TYR456 MET542 |
| **2** | -5.99 | 40.96 | ALA348 HIS401 ARG514 GLU402 | -6.46 | 18.32 | ASP623 THR556 ARG624 THR680 SER681 SER682 | -2.88 | 7.81 | ARG553 ASP623 ASN691 |

**A Chain:** ARG858 ILE548 ALA547 ARG836

| | | | | | | MET542 VAL667 ALA558 | | | |
|---|---|---|---|---|---|---|---|---|---|
| 3 | -5.70 | 66.75 | ALA348 ASP350 ASP382 HIS378 HIS401 | -6.35 | 22.18 | ASP623 ARG624 MET542 ALA558 VAL557 THR680 | -2.88 | 7.71 | LYS621 ASP623 SER682 THR556 CYS622 TYR619 |
| 4 | -5.58 | 81.21 | ALA348 ASP382 HIS401 HIS374 HIS378 | -6.22 | 27.52 | TYR619 ASP623 SER681 SER682 ALA558 MET542 ARG624 | -2.68 | 10.82 | THR680 ASN691 ASP623 ARG553 TYR455 LYS621 TYR619 CYS622 |
| 5 | -5.48 | 95.81 | ALA348 ASP350 ASP382 HIS401 HIS374 Zn704 GLU402 HIS378 | -6.17 | 30.12 | ASP623 ALA554 THR556 SER682 SER681 ARG624 MET542 ALA558 | -2.67 | 11.05 | TYR619 LYS621 ARG624 ALA558 TYR456 MET542 ASP623 |
| 6 | -5.14 | 171.38 | ASP382 HIS401 ALA348 ASP350 | -6.01 | 39.46 | ASP452 ARG624 ASP623 ALA558 VAL557 MET542 VAL667 SER681 TYR456 SER682 | -2.42 | 16.97 | ASP623 CYS622 |
| 7 | -5.05 | 198.05 | GLU398 ARG514 GLU402 Zn704 THR347 ALA348 HIS378 HIS401 | -5.92 | 46.01 | ASN691 THR680 ASP760 ALA558 MET542 VAL557 ARG624 SER682 ASP623 | -1.92 | 39.02 | ASP623 TYR455 |
| 8 | -5.05 | 197.39 | ALA348 HIS378 | -5.27 | 135.97 | ARG553 ALA554 ASP452 ASP623 TYR456 SER681 SER682 ARG624 ALA558 MET542 VAL667 | -1.26 | 118.92 | SER682 ASP623 ARG553 LYS621 TYR455 |
| 9 | -4.83 | 286.00 | GLU375 THR347 ALA348 ASP382 ARG393 PHE390 ASP350 | -5.16 | 164.39 | CYS622 ASP623 ASP452 ARG624 LYS621 TYR455 ARG553 | -0.74 | 286.08 | SER682 ARG553 ARG624 ASP623 LYS621 TYR455 |
| 10 | -4.47 | 526.55 | ASP382 ALA348 ASP350 TRP349 HIS401 | -4.78 | 315.86 | ALA554 ASP623 ARG624 SER681 SER682 MET542 VAL667 ALA558 | -0.55 | 393.12 | CYS622 ASP623 ARG553 LYS621 ARG624 |

**4. The binding mode analysis and predicated binding affinity calculations of hydroxychloroquine (HCQ) against the homology model of COVID-19 RNA-dependent RNA polymerase (RdRP)**:  As mention in the binding mode of CQ with RdRp, Therefore, this web-based server was utilized for docking studies of HCQ. The docking studies of RdRp with RNA revealed the interaction of **HCQ** with A Chain:

**A Chain:** ARG858 ILE548 ALA547 ARG836

**P Chain: G6** G7 as shown in **figure 5 A**. The negative values of the binding free energy (-7.20 kcal/mole) further indicated the stability of the complex (**Table 2**). The receptor pocket and top 10 conformers are shown in **figure 5 A**. Similarly, the docking studies of HCQ on RdRp without RNA revealed the interaction of HCQ with **A Chain: TYR456** THR680 TYR455 ARG553 ASP623 as shown in **figure 5 B**. The negative values of the binding free energy (-6.00 kcal/mole) further indicated the stability of the complex (**Table 2**). The receptor pocket and top 10 conformers are shown in **figure 5 A.**

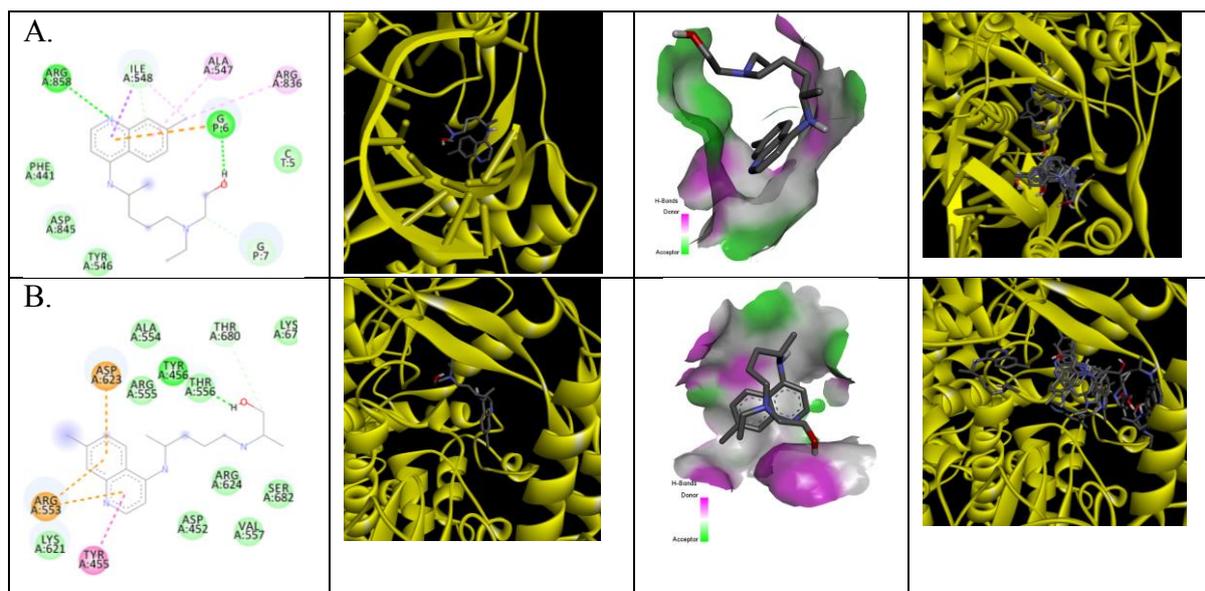

**Figure 5**: Best Conformation observed due to the binding of ligand hydroxychloroquine with Nonstructural protein 12 (nsp12) of SARS-CoV; **A:** nsp 12 RdRp with RNA (Note: There are 2 neighboring binding pockets) and **B:** nsp 12 RdRp without RNA (Note: There are 2 neighboring binding pockets)

**Note**: Highlighted portion with yellow color indicates the amino acids interact with -OH group present in **HCQ**.

**5. The binding mode analysis and predicated binding affinity calculations of remdesivir against surface receptor and nonstructural proteins**: Molecular docking studies of remdesivir (RDV) against the novel coronavirus spike receptor-binding domain complexed with its receptor ACE2 (PDB ID 6LZG) revealed the interaction of RDV with TYR385 ALA348 GLU398 ARG514 as shown in **figure 6 A**. The negative values of the binding free energy (-4.55 kcal/mole) further indicated the stability of the complex (**Table 5**). The receptor pocket and top 10 conformers are shown in **figure 6 A**. Similarly, the docking studies and binding mode analysis of RDV against SARS-Coronavirus NSP12 bound to NSP7 and NSP8 co-factors (**PDB ID 6NUR**) showed the interactions with ALA558 LYS676 ARG553 ASP623 ARG555 SER682 TYR456 VAL667 VAL557 ARG624 (**figure 6 B**). The negative free energy calculation (-4.34 kcal/mole) as shown in table 3 is an indication of the interaction of RDV with nonstructural viral proteins. Our docking studies of **RDV** with NSP16-NSP10, however, our docking methodology did not work very well with this target resulted in the high positive values of binding energy (data are not shown here). We will further improve our next publications.

**Table 4:** Binding affinity of Hydroxychloroquine ($C_{18}H_{26}OClN_3$) with the target Nonstructural protein 12 (nsp12) of SARS-CoV

| Cluster Rank | NSP 12 RdRp with RNA | | NSP 12 RdRp without RNA | |
|---|---|---|---|---|
| | **Energy of binding** | **Interacting Amino Acids** | **Energy of binding** | **Interacting Amino Acids** |
| 1 | -7.20 | **A Chain:** ARG858 ILE548 ALA547 ARG836 <br> **P Chain:** G6 G7 | -6.00 | **A Chain:** TYR456 THR680 TYR455 ARG553 ASP623 |
| 2 | -7.00 | **A Chain:** ARG836 ILE548 ALA547 ARG858 PHE441 <br> **P Chain:** G6 | -5.90 | **A Chain:** ASP623 ARG624 MET542 ALA558 THR556 |
| 3 | -6.90 | **A Chain:** ARG858 ALA840 ILE548 <br> **P Chain:** G6 G7 | -5.70 | **A Chain:** ALA688 ALA685 LYS500 ARG569 TYR689 |
| 4 | -6.70 | **A Chain:** ALA840 ILE548 ARG836 ARG555 <br> **P Chain:** G6 G7 | -5.70 | **A Chain:** ARG555 LYS545 ARG858 ILE548 ARG836 ALA547 |
| 5 | -6.60 | **A Chain:** ARG836 ILE548 ARG858 ALA547 PHE441 ALA840 ARG555 <br> **P Chain:** G7 | -5.60 | **A Chain:** ASP452 THR556 ASP623 SER682 ARG624 ALA558 |
| 6 | -6.50 | **A Chain:** ILE548 ARG555 ASP845 <br> **P Chain:** C5 G7 G8 | -5.40 | **A Chain:** ARG858 ILE548 ALA840 ARG836 |
| 7 | -6.40 | **A Chain:** ARG858 ILE548 ARG836 ALA840 <br> **P Chain:** C5 G7 | -5.40 | **A Chain:** ARG553 ILE548 ARG858 PHE441 ALA840 ALA547 ARG555 |
| 8 | -6.40 | **A Chain:** LYS545 THR556 ASP623 ALA558 MET542 VAL667 ARG624 <br> **P Chain:** G8 <br> **T Chain:** G2 | -5.40 | **A Chain:** ASP623 LYS621 TYR455 ARG553 |
| 9 | -6.20 | **A Chain:** THR556 SER682 ASP623 <br> **P Chain:** G8 | -5.40 | **A Chain:** SER682 THR556 ASP623 |
| 10 | -6.20 | **P Chain:** G6 G7 C5 | -5.30 | **A Chain:** SER682 THR556 ASP760 ASP623 CYS622 |

**A.**

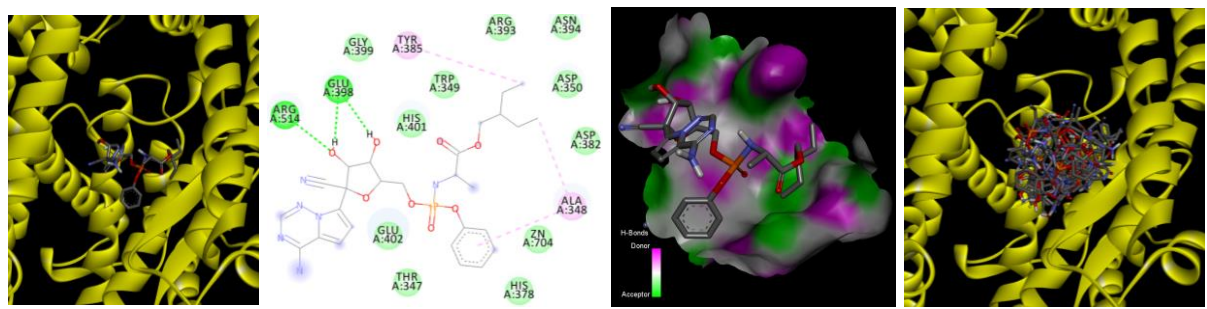

**B.**

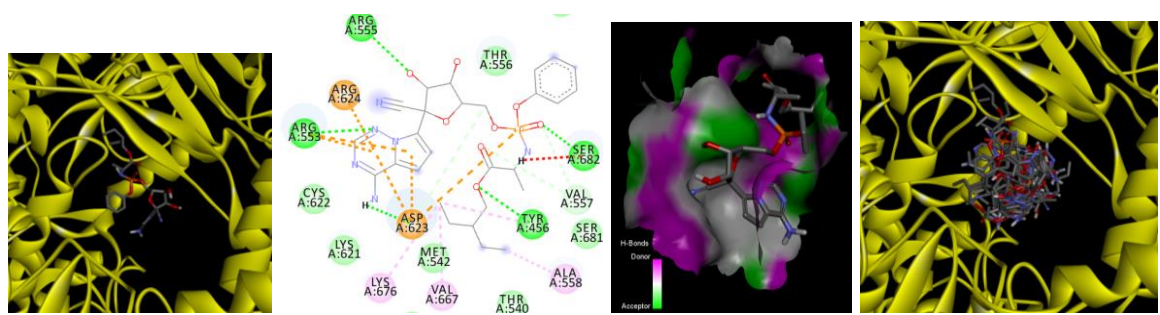

**Figure 6**: Conformational changes observed due to the binding of ligand remdesivir with **A.** PDB ID: 6LZG, **B**. PDB ID: 6NUR; left to right there are the best pose, interaction 2D image of best pose, receptor pocket, and top10 conformers.

**Table 5**: Binding affinity of remdesivir ($C_{27}H_{35}N_6O_8P$) with the target protein PDB ID: 6LZG, 6NUR

| Cluster Rank | PDB ID: 6LZG | | | PDB ID: 6NUR | | |
|---|---|---|---|---|---|---|
| | Free Energy of Binding (kcal/mol) | Predicted Inhibition Constant (µM) | Interacting Amino Acids | Free Energy of Binding (kcal/mol) | Predicted Inhibition Constant (µM) | Interacting Amino Acids |
| 1 | -4.55 | 461.75 µM | TYR385 ALA348 GLU398 ALA514 | -4.34 | 661.90 µM | ALA558 LYS676 ARG553 ASP623 ARG555 SER682 TYR456 VAL667 VAL557 ARG624 |
| 2 | -4.32 | 679.89 µM | HIS378 HIS401 ASN394 ALA348 TRP349 ASP350 | -4.10 | 995.09 µM | ASP452 ALA554 ARG624 ARG555 THR556 MET542 ALA558 SER682 TYR456 LYS621 ARG553 ASP623 |
| 3 | -3.92 | 1.34 mM | GLU402 ALA348 ASN394 THR347 ASP382 HIS401 | -3.82 | 1.59 mM | TYR455 ARG624 CYS622 LYS621 ARG555 ASP623 ASP452 ALA554 THR556 ARG553 |
| 4 | -3.44 | 2.99 mM | ARG514 HIS401 GLU402 ASP382 ALA348 HIS345 HIS378 GLU398 | -3.60 | 2.29 mM | LYS621 TYR455 ARG553 ARG624 THR556 ALA554 ASP623 CYS622 |
| 5 | -3.09 | 5.42 mM | Zn704 ALA348 ASP382 HIS378 HIS401 ARG514 | -3.50 | 2.70 mM | LYS545 ARG555 ASP623 CYS622 ASP760 ARG624 ARG553 SER682 VAL557 |
| 6 | -2.90 | 7.48 mM | ARG393 PHE40 ASN394 ALA348 ASN401 ASP382 Zn704 GLU402 HIS378 | -2.95 | 6.92 mM | ASP760 TYR455 CYS622 ASP452 ASP623 THR556 ARG624 ARG555 ARG553 LYS621 |
| 7 | -2.81 | 8.75 mM | HIS345 THR347 TRP349 ALA348 HIS401 HIS378 ASP382 | -1.08 | 160.23 mM | LYS621 ARG624 THR556 ASP452 ASP623 CYS622 MET626 ARG553 |
| 8 | -2.78 | 9.15 mM | TRP349 ASP350 ALA348 HIS378 ASP382 ARG514 GLU398 ASN397 PHE400 | -0.62 | 353.02 mM | THR556 TYR455 ARG624 THR680 ASP623 ALA554 ARG553 LYS621 ARG555 |
| 9 | -2.45 | 15.87 mM | ASP350 GLU375 PRO346 HIS345 HIS378 GLU402 HIS401 ALA348 | -0.44 | 472.51 mM | ARG624 LYS621 ARG555 THR556 TYR619 ARG553 ASP452 ASP623 |
| 10 | -2.18 | 25.03 mM | HIS345 HIS401 ARG393 ALA348 PHE40 THR347 ASP350 | 4.23 | NA | TYR619 ASP623 ARG555 ALA554 ASP452 ARG624 THR556 ARG553 LYS621 |

**6. The binding mode analysis and predicated binding affinity calculations of remdesivir against the homology model of COVID-19 RNA-dependent RNA polymerase (RdRP)**:   As mentioned earlier the web-based was utilized for docking studies of RDV. The docking studies of RdRp with RNA revealed the interaction of RDV with A Chain: **A Chain:** PHE441 ALA547 ILE548 LYS545 ARG624 ASP452

ASP623 ARG555 **T Chain:** G2 **P Chain:** G7 G8 as shown in **figure 7 A**. The negative values of the binding free energy (-9.40 kcal/mole) further indicated the stability of the complex (**Table 6**). The receptor pocket and top 10 conformers are shown in **figure 7 A**. Similarly, the docking studies of RDV on RdRp without RNA revealed the interaction of RDV **A Chain:** TYR455 ARG553 ASP760 CYS622 ASP623 ARG624 THR680 TYR456 ALA558 THR556 as shown in **figure 7 A**. The negative values of the binding free energy (-8.50 kcal/mole) further indicated the stability of the complex (**Table 6**). The receptor pocket and top 10 conformers are shown in **figure 6 A.**

A.

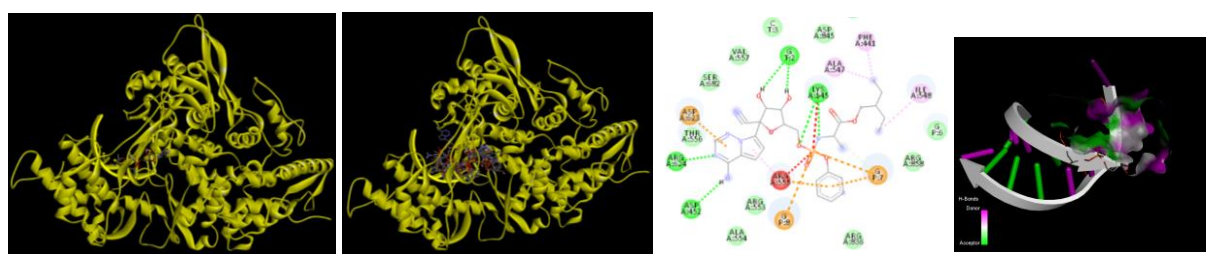

B.

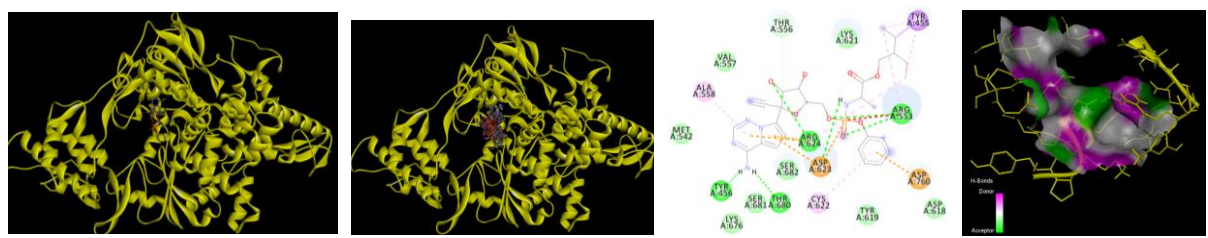

**Figure 7**: Best Conformation observed due to the binding of ligand remdesivir with Nonstructural protein 12 (nsp12) of SARS-CoV: **A**. nsp 12 RdRp with RNA and **B**. nsp 12 RdRp without RNA

**Table 6**: Binding affinity of remdesivir ($C_{27}H_{35}N_6O_8P$) with the target nonstructural protein 12 (nsp12) of SARS-CoV

| Cluster Rank | NSP12 RdRp with RNA | | NSP 12 RdRp without RNA | |
|---|---|---|---|---|
| | Energy of binding | Interacting Amino Acids | Energy of binding | Interacting Amino Acids |
| 1 | -9.40 | **A Chain:** PHE441 ALA547 ILE548 LYS545 ARG624 ASP452 ASP623 ARG555 **T Chain:** G2 **P Chain:** G7 G8 | -8.50 | **A Chain:** TYR455 ARG553 ASP760 CYS622 ASP623 ARG624 THR680 TYR456 ALA558 THR556 |
| 2 | -9.40 | **A Chain:** LYS545 THR556 ARG553 LYS621 TYR455 ARG624 ASP623 ASP760 **T Chain:** G2 **P Chain:** G8 | -8.30 | **A Chain:** TYR455 LYS621 ARG553 THR556 ASP623 ARG624 ALA558 TYR456 THR680 CYS622 ASP760 |

| 3 | -9.30 | **A Chain:** ARG555 ALA547 PHE441 ILE548 <br> **P Chain:** C4 C5 G6 G7 G8 | -8.10 | **A Chain:** ASP760 THR556 ARG624 ASP452 TYR455 ARG553 ASP623 |
|---|---|---|---|---|
| 4 | -9.10 | **A Chain:** SER682 LYS545 ARG555 ILE548 ASP760 <br> **P Chain:** G8 | -7.90 | **A Chain:** THR680 ARG624 TYR455 LYS621 ARG553 THR556 SER682 ASP623 |
| 5 | -9.10 | **A Chain:** VAL557 LYS545 ARG555 ALA547 ILE548 ARG858 ALA840 ASP845 <br> **P Chain:** C5 G6 G7 | -7.70 | **A Chain:** ASP760 LYS621 CYS622 THR680 TYR456 ALA558 ARG624 ASP623 THR556 ARG553 |
| 6 | -9.00 | **A Chain:** ALA840 ILE548 LYS545 ARG555 ALA547 PHE441 ASP452 ASP623 <br> **P Chain:** G8 <br> **T Chain:** G2 | -7.70 | **A Chain:** LYS551 LYS621 TYR455 THR680 ASP623 ARG553 THR556 ARG555 SER682 |
| 7 | -9.00 | **A Chain:** ILE548 PHE441 ASP845 ASN497 LYS500 ASP499 LYS545 ARG555 <br> **T Chain:** C3 G2 <br> **P Chain:** G7 | -7.70 | **A Chain:** ASP618 TYR455 LYS621 ARG553 ARG624 ASP623 THR556 SER682 ASN691 THR680 |
| 8 | -9.00 | **A Chain:** ARG555 LYS545 ILE548 ALA840 ARG836 ASP623 ARG624 ASP452 <br> **T Chain:** G2 <br> **P Chain:** G7 G8 | -7.60 | **A Chain:** ASP452 TYR455 LYS621 ARG553 ARG624 ASP623 THR680 ASP760 THR556 |
| 9 | -9.00 | **A Chain:** SER682 ASP623 ARG624 ARG555 ALA840 ALA547 PHE441 ILE548 LYS545 <br> **T Chain:** G2 C3 | -7.50 | **A Chain:** TYR455 ARG553 LYS621 ASP760 CYS622 ASP623 |
| 10 | -8.90 | **A Chain:** ASP452 ARG553 TYR455 ASP623 CYS622 ASP760 <br> **P Chain:** G8 <br> **T Chain:** G2 | -7.50 | **A Chain:** TYR455 ASP452 ARG553 ASP760 THR680 ALA688 ASP623 CYS622 LYS621 |

**7. The binding mode analysis and predicated binding affinity calculations of arbidol against surface receptor and nonstructural proteins**: Molecular docking studies of **arbidol** (ABD) against the novel coronavirus spike receptor-binding domain complexed with its receptor ACE2 (PDB ID 6LZG) revealed the interaction of ABD with **ASP350 TRP349 ALA348 HIS378 ASP382 HIS401** as shown in **figure 8A**. The negative values of the binding free energy (-7.49 kcal/mole) further indicated the stability of the complex (**Table 7**). The receptor pocket and top 10 conformers are shown in **figure 8A**. Similarly, the docking studies and binding mode analysis of ABD against SARS-Coronavirus NSP12 bound to NSP7 and NSP8 co-factors (**PDB ID 6NUR**) showed the interactions with **ASP452 ARG553 THR556 ASP623 ARG624** (**Figure 8B**). The negative free energy calculation (-6.31 kcal/mole) as shown in table 5 is an indication of the interaction of ABD with nonstructural viral proteins. Our docking studies of ABD with NSP16- NSP10, however, our docking methodology did not work

very well with this target resulted in the high positive values of binding energy (data are not shown here). We will further improve our next publications.

**A.**

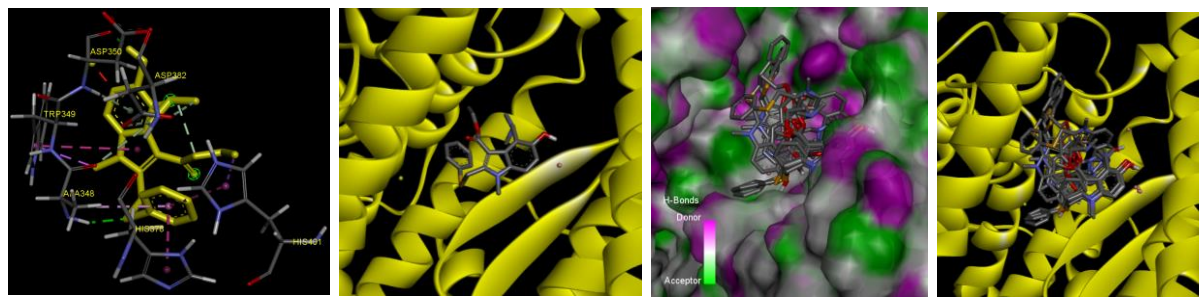

**B.**

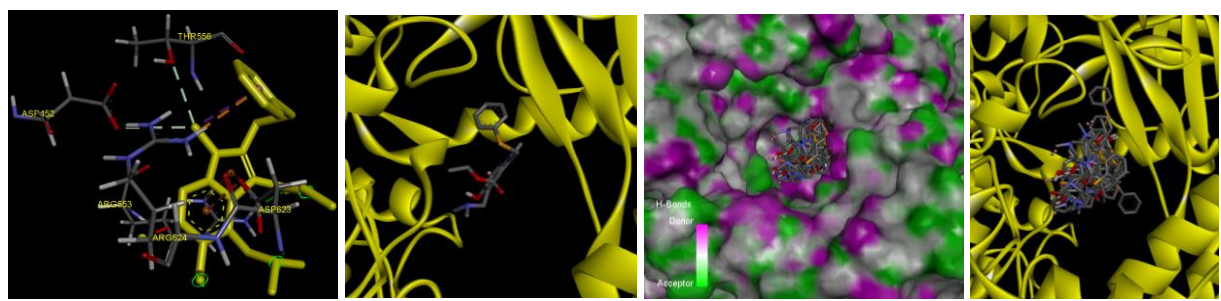

**Figure 8**: Conformational changes observed due to the binding of ligand Arbidol with A. PDB ID: 6LZG, B. PDB ID: 6NUR; left to right best pose, receptor pocket, and top10 conformers.

**Table 7**: Binding affinity of arbidol ($C_{22}H_{25}N_2SO_3Br$) with the target protein PDB ID: 6LZG, 6NUR.

| Cluster Rank | PDB ID: 6LZG | | | PDB ID: 6NUR | | |
|---|---|---|---|---|---|---|
| | Free Energy of Binding (kcal/mol) | Predicted Inhibition Constant (μM) | Interacting Amino Acids | Free Energy of Binding (kcal/mol) | Predicted Inhibition Constant (μM) | Interacting Amino Acids |
| 1 | -7.94 | 1.51 | ASP350 TRP349 ALA348 HIS378 ASP382 HIS401 | -6.31 | 23.79 μM | ASP452 ARG553 THR556 ASP623 ARG624 |
| 2 | -7.68 | 2.36 | ASP382 HIS401 ASP350 TRP349 ALA348 HIS378 | -5.69 | 67.69 μM | TYR456 MET542 THR556 ALA558 ASP623 ARG624 VAL667 THR680 |
| 3 | -7.61 | 2.65 | ASP350 TRP349 ALA348 GLU375 ASP382 | -5.36 | 117.68 μM | ARG553 ARG555 THR556 ASP623 ARG624 SER682 ASP760 |
| 4 | -7.28 | 4.64 | ALA348 ASP382 ASN394 HIS401 GLU402 | -5.04 | 200.60 μM | TYR455 ARG553 ARG555 THR556 ASP623 ARG624 |
| 5 | -7.09 | 6.37 | ASP350 TRP349 ALA348 HIS378 ASP382 HIS401 | -4.65 | 388.54 μM | ASP452 ARG553 ALA554 THR556 CYS622 ASP623 ARG624 |

| 6 | -7.01 | 7.22 | HIS378 ASP382 ASN394 HIS401 GLU402 | -4.52 | 489.35 µM | ASP452 LYS545 ARG553 ALA554 ARG555 THR556 ASP623 ARG624 |
| 7 | -6.72 | 11.96 | ALA348 ASP350 HIS378 ASP382 HIS401 ARG514 | -4.48 | 520.82 µM | ARG553 ARG555 THR556 ASP623 ASP760 |
| 8 | -6.72 | 11.89 | GLU375 HIS378 ASP382 HIS401 | -3.93 | 1.32 mM | ARG553 THR556 ALA558 ASP623 ARG624 ASP760 |
| 9 | -6.39 | 20.68 | ASP350 TRP349 ALA348 HIS378 ASP382 HIS401 | -3.51 | 2.69 mM | ASP452 ARG553 THR556 ASP623 |
| 10 | -5.92 | 45.87 | ALA348 HIS378 ASP382 ASN394 HIS401 GLU402 | -3.48 | 2.82 mM | TYR455 ASP623 ARG624 ASP760 |

**8. The binding mode analysis and predicated binding affinity calculations of arbidol against the homology model of COVID-19 RNA-dependent RNA polymerase (RdRP):** As mentioned earlier the web-based was utilized for docking studies of ABD. The docking studies of RdRp with RNA revealed the interaction of ABD with TYR456 THR680 ASP452 ARG624 **figure 9 A**. The negative values of the binding free energy (-7.30 kcal/mole) further indicated the stability of the complex (**Table 8**). The receptor pocket and top 10 conformers are shown in **figure 9 A**. Similarly, the docking studies of ABD on RdRp without RNA revealed the interaction of ABD **A Chain:** TYR455 ARG553 ASP760 CYS622 ASP623 ARG624 THR680 TYR456 ALA558 THR556 as shown in **figure 9 A**. The negative values of the binding free energy (-5.90 kcal/mole) further indicated the stability of the complex (**Table 8**). The receptor pocket and top 10 conformers are shown in **figure 9 A.**

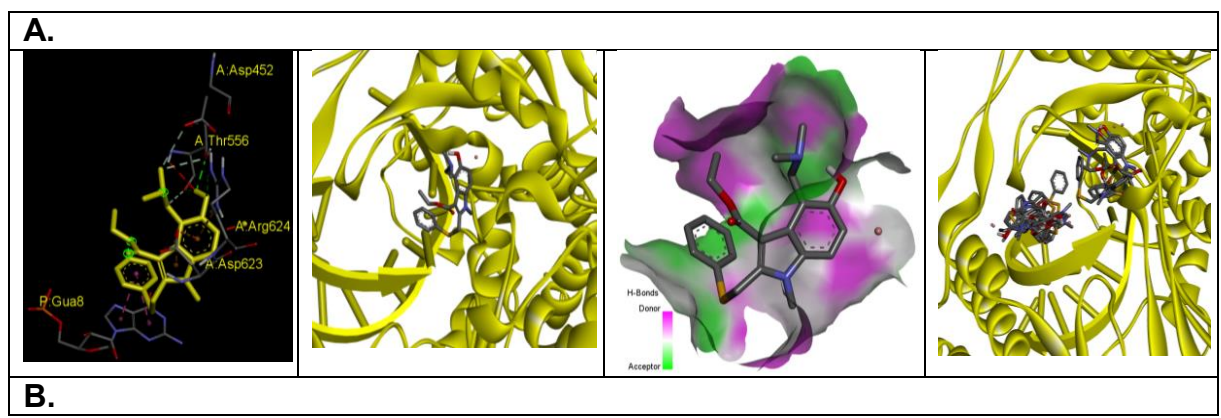

**A.**

**B.**

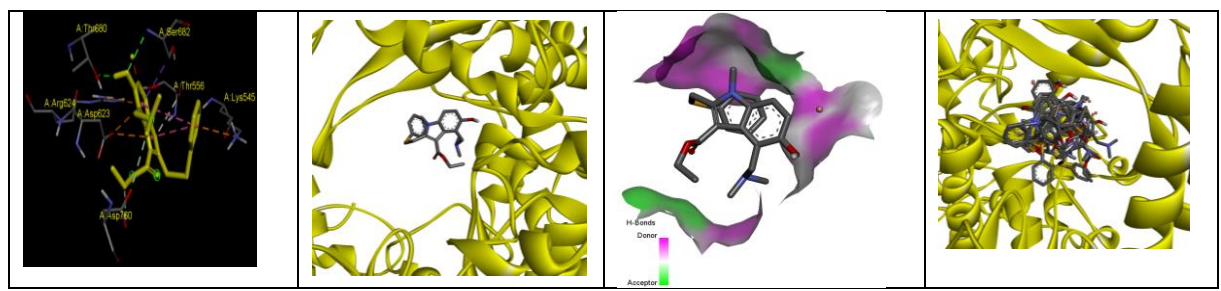

**Figure 9**: Best Conformation observed due to the binding of ligand arbidol with Nonstructural protein 12 (nsp12) of SARS-CoV; **A:** nsp 12 RdRp with RNA and **B:** nsp 12 RdRp without RNA.

**Table 8**: Binding affinity of arbidol ($C_{22}H_{25}N_2SO_3Br$) with the target nonstructural protein 12 (nsp12) of SARS-CoV

| Cluster Rank | NSP 12 RdRp with RNA | | NSP 12 RdRp without RNA | |
|---|---|---|---|---|
| | **Energy of binding** | **Interacting Amino Acids** | **Energy of binding** | **Interacting Amino Acids** |
| 1 | -7.30 | **A Chain:** ASP452 THR556 ASP623 ARG624 **P Chain:** G8 | -5.90 | **A Chain:** LYS545 THR556 ASP623 ARG624 THR680 SER682 ASP760 |
| 2 | -7.10 | **A Chain:** PHE441 LYS545 ALA547 ILE548 ARG555 ASP845 ASP858 **P Chain:** G6 G7 | -5.70 | **A Chain:** ASP452 THR556 ASP623 ARG624 ALA688 |
| 3 | -7.10 | **A Chain:** ILE548 ARG555 ASP845 **P Chain:** G7 G8 | -5.70 | **A Chain:** ASP452 ARG553 THR566 ALA558 ASP623 ARG624 |
| 4 | -7.00 | **A Chain:** LYS545 ILE548 ARG555 **P Chain:** C5 G6 G7 | -5.30 | **A Chain:** ASP452 ARG553 THR556 TYR619 ASP623 |
| 5 | -7.00 | **A Chain:** PHE441 LYS545 ALA547 ILE548 ARG555 VAL557 ASP845 ARG858 **P Chain:** G6 G7 | -5.30 | **A Chain:** ASP452 ALA554 THR556 ASP623 ARG624 ASP760 |
| 6 | -7.00 | **A Chain:** LYS545 ARG555 **P Chain:** C5 G6 G7 | -5.30 | **A Chain:** ASP452 ALA554 THR556 ASP623 ARG624 |
| 7 | -6.80 | **A Chain:** THR556 ASP623 ARG624 **P Chain:** G8 | -5.20 | **A Chain:** THR556 ASP623 ARG624 ALA688 |
| 8 | -6.80 | **A Chain:** LYS545 ARG555 **P Chain:** G7 G8 | -5.20 | **A Chain:** THR556 ASP618 ASP623 ASP760 |
| 9 | -6.70 | **A Chain:** ARG555 ASP623 ASP760 **P Chain:** G8 | -5.10 | **A Chain:** LYS545 ARG553 ARG555 THR556 ASP623 |
| 10 | -6.70 | **A Chain:** LYS545 **P Chain:** C5 G6 G7 G8 **T Chain:** C4 | -5.10 | **A Chain:** ARG553 CYS622 ASP623 ASP760 |

**Conclusion**

Here, we report the preliminary docking studies of the selected FDA approved drugs which have shown promising anti-COVID-19 activity. Literature evidences are indicating a single drug is probably not effective for the treatment of this disease. The viral polymerases and proteases are known to be the most suitable targets for the treatment of viral born disease. Therefore, we selected the targets to cover a range of targets starting from the entry point to the multiplication and release of the viruses. In our *in-silico* experiment based on mainly molecular docking approach, we investigated four different types of pharmacologically active and US-FDA approved drugs towards their potential application alone or in combination with drug repurposing. Our studies can give an insight into the possible the pocket(s) of ligand interaction, best conformer and its orientation and interactions with the respective targets. The molecules are arranged in decreasing order of their binding free energies, in respective tables. Based on their binding affinity, we found that these compounds may be potential inhibitors against SARS-CoV having promising interactions with the selected targets.

**Future direction and limitations of the study**: The work presented here is a preliminary study based on the hypothesis to repurpose the FDA approved drugs to identify the combination therapy potentially targeting the various stages of the virus life cycle for the management of COVID-19. The study needs further investigation through the various computational tool and further detailed biological evaluation for the selection of possible combination therapy to potentially deal with the current pandemic situation.

**Acknowledgment**

GM is thankful to the Indian Institute of Technology (BHU). AB is thankful to Indian Institute of Technology (BHU) and MHRD, India for fellowship. AB is thankful to Gyan Modi for giving such an opportunity and provides all kinds of supports to complete this work-from-home initiative.